\documentstyle[12pt,epsf]{article}

\begin{document}

\title{
{E-print hep-ph/9808363 \hfill Preprint YARU-HE-98/06} \\[10mm]
\bf Mass effects in the quark-gluon decays \\ 
            of heavy paraquarkonia }

\author{A.Ya.~Parkhomenko and A.D.~Smirnov\thanks{
        E-mail: asmirnov@univ.uniyar.ac.ru}\\
{\small\it Department of Theoretical Physics, Yaroslavl State University,}\\
{\small\it Sovietskaya 14, 150000 Yaroslavl, Russia.}}

\date{}

\maketitle

\begin{abstract}
Quark-gluon decays of heavy paraquarkonia 
$^1 S_0(\bar Q Q) \to \bar q q g$ 
are investigated with account of the masses of final quarks. 
The decay widths and the energy distributions 
of the final quarks and gluons 
are calculated in dependence on the relative quark masses. 
The strong collinear enhancement 
of the gluon energy distribution 
at the end of the spectrum 
is shown to take place 
in all such decays of 
$\eta_c$ and $\eta_b$ 
mesons except  the decay 
$\eta_b \to \bar c c g$.
The total decay width is shown to have 
an essential dependence
on the final quark masses.  
The corresponding branching ratios 
of $\eta_c$ and $\eta_b$ mesons 
are numerically estimated with a good agreement of 
$Br(\eta_c \to \bar s s g)$ 
with experimental data on $\eta_c$ decays.

\end{abstract}

\vspace*{10mm}

\centerline{{\it Mod. Phys. Lett.}, 1998, V. {\bf A13}, No. 23, P.} 

\thispagestyle{empty}

\newpage 

Many-partical decays of heavy quarkonia such as  
$^1 S_0 (\bar Q Q)\to 3g, \bar q q g$,  
$^3 S_1 (\bar Q Q)\to 4g, \bar q q g g$
depend on the $3g$- and $\bar q q g$-interactions of the final 
quarks and gluons and can give the useful information about 
these interactions. In particular, these interactions give rise 
to the specific thrust~\cite{Barb1,Kamal} 
and invariant masses~\cite{Koller,Streng} distributions, 
to the acomplanarity of the four-particle decays~\cite{Muta}, 
to the collinearization effect of final gluons~\cite{AD1}, etc. 
It should be noted, however, that such effects 
can be essentially affected by  masses of final quarks. 
For example, the four-particle quark-gluon decays of 
the orthobottomonium with the production of $\bar u u$-, $\bar d d$- or 
$\bar s s$-pairs exhibit a collinear enhancement whereas this effect 
in such decays with the production of $\bar c c$-pair is absent 
because of the relatively large mass of $c$-quark~\cite{AD2}. 
Hence, the effect of the final quark masses can be essential 
in the decays of heavy quarkonia with production of $\bar q q$-pairs 
and it should be taken into account in investigations of such decays. 

In this letter the quark-gluon decay of heavy paraquarkonium 
$^1 S_0(\bar Q Q) \to \bar q q g$ 
is investigated taking account of the masses of final quarks. 
The differential and total decay widths are calculated in tree 
approximation in dependence on the final quark masses and are 
discussed in comparison with the experimental data on $\eta_c$ decays. 
The branching ratios of $\eta_b$ decay are also predicted and discussed. 
  
\begin{figure}[tb]
\begin{center}
\centerline{\epsfxsize=\textwidth \epsffile[110 565 520 675]{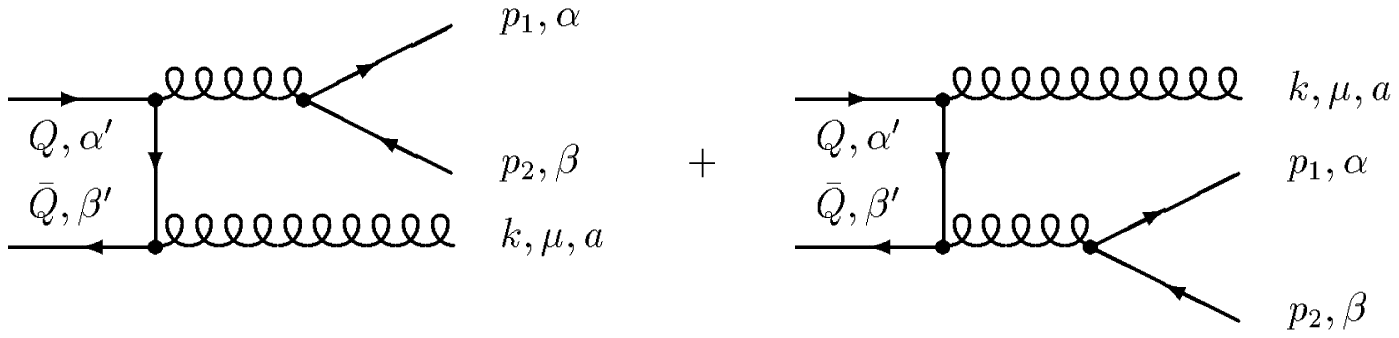}}
\vfill
A.Ya.~Parkhomenko and A.D.Smirnov \\ 
Mass effects in the quark-gluon decays of heavy paraquarkonia \\ 
\end{center}
\caption{}
\label{fig:QQqqG}
\end{figure}

 The decay of a heavy paraquarkonium into a quark-antiquark 
 pair and a gluon $^1S_0 (\bar Q Q) \to \bar q q g$ 
 is described by two graphs shown in Fig.~\ref{fig:QQqqG}.
 The amplitude of this process in the limit of static quarks
 in the quarkonium can be presented in the form:

\begin{equation}
{\cal M}^a_{\alpha \beta} (^1S_0 \to \bar q q g) = -
\frac{(t_a)_{\alpha \beta}}{2 \sqrt N_c} \,
\frac{g_s^3 \psi (0)}{2 m \sqrt{\omega \varepsilon_1 \varepsilon_2}} \,
\frac{(j_{\alpha \beta} \tilde F^a {\cal P})}{p^2 \, ({\cal P} k)}, 
\label{eq:amplinv} 
\end{equation}

 \noindent 
 where $t_a$ are the generators of the color group $SU(N_c)$, 
 $a= 1, 2, \ldots, N_c^2 - 1$ and $\alpha, \beta = 1, 2, \ldots, N_c$ 
 are the color indices, $g_s$ is the strong charge,
 $\psi ({\bf r})$ is the nonrelativistic wave function of a paraquarkonium 
 in the coordinate space, 
 $\omega$, $\varepsilon_1$ and $\varepsilon_2$ are the energies of the
 gluon, quark and antiquark respectively,
 $(j_\mu)_{\alpha \beta} = (\bar u_\alpha (p_1) \gamma_\mu u_\beta (- p_2))$ 
 is the final quark current,
 $\tilde F^a_{\mu \nu} = \varepsilon_{\mu \nu \rho \sigma} k_\rho e^a_\sigma$,
 $e^a_\mu$ and $k_\mu$ are the polarization and wave vectors of the gluon,
 $p = p_1 + p_2$, $p_{1 \mu}$ and $p_{2 \mu}$ are four-momenta of 
 the final quark and antiquark,
 ${\cal P}_\mu \simeq (2 m, {\bf 0})$ is the four-momentum of a 
 paraquarkonium in its rest frame, and 
 $m$ is the mass of the heavy quark in the quarkonium.

 The differential probability of the decay after summation over colors 
 and polarizations of the final particles can be written in the form:

\begin{eqnarray}
d\Gamma_{\bar q q g} & = & F_c \, 
\frac{\alpha_s^3 | \psi(0) |^2}{\pi^2 m^2} \,
\left [ 
\frac{({\cal P} (p_1 - p_2))^2 + ({\cal P} k)^2}{p^2 ({\cal P} k)^2} +
\frac{4 m^2 \mu^2}{(p^2)^2}
\right ] 
\nonumber \\
& \times &
\delta^{(4)} ({\cal P} - p_1 - p_2 - k) \,
\frac{d{\bf p_1} d{\bf p_2} d{\bf k}}{\varepsilon_1 \varepsilon_2 \omega} ,
\label{eq:diffw}
\end{eqnarray}

\noindent 
 where $F_c = (N_c^2 - 1)/(8 N_c)$ is the color factor 
 of the $SU (N_c)$ group
 (in QCD, $N_c = 3$ and $F_c = 1/3$), 
 $\alpha_s = g_s^2/(4 \pi)$ is the strong coupling constant,
 and $\mu = m_q /m$ is the relative mass of a final quark (antiquark). 

 The probability~(\ref{eq:diffw}) of the three-particle decay
 $^1S_0 (\bar Q Q) \to \bar q q g$ of the para\-quar\-ko\-nium 
 depends on two independent variables and 
 as a function of two independent energies 
 can be presented in the form:

\begin{equation}
\frac{d^2\Gamma_{\bar q q g}}{dy_1 dy_2} = 
\frac{d^2\Gamma_{\bar q q g}}{dy_1 dx} = F_c \, 
\frac{2 \alpha_s^3 | \psi(0) |^2}{m^2} \,
\left [ 
\frac{x^2 + (y_1 - y_2)^2}{x^2 (1 - x)} + \frac{\mu^2}{(1 - x)^2}
\right ] ,
\label{eq:diffwen}
\end{equation}

\noindent 
 where $y_1 = \varepsilon_1 / m$, $y_2 = \varepsilon_2 / m$ and
 $x = \omega / m$ are the relative energies of the quark, antiquark 
 and gluon respectively satisfying the energy conservation 
 low $x + y_1 + y_2 = 2$.
 The expression~(\ref{eq:diffwen}) is rather simple 
 and allows for the relative mass of the final quark. 
 In the particular case of $\mu = 0$ it is consistent 
 with the corresponding result presented in the more 
 complicated form in Ref.~\cite{Kamal}. 

 Integrating Eq.~(\ref{eq:diffw}) over the momenta of the quark 
 ${\bf p_1}$ and the antiquark ${\bf p_2}$ or over the momenta of
 the antiquark ${\bf p_2}$ and the gluon ${\bf k}$, we obtain the 
 distribution in the energy $x$ of the final gluon or that in the 
 energy $y_1$ of the final quark in the form:

\begin{equation}
\frac{d \Gamma_{\bar q q g}}{d x} =
F_c \, \frac{8 \alpha_s^3 | \psi(0) |^2}{3 m^2} \, 
\frac{x}{1 - x} \,
\left (
1 + \frac{\mu^2}{2 (1 - x)}
\right ) \,
\sqrt{1 - \frac{\mu^2}{1 - x}} ,
\label{eq:distrx} 
\end{equation}

\begin{eqnarray}
\frac{d \Gamma_{\bar q q g}}{d y_1} & = & 
F_c \, \frac{4 \alpha_s^3 | \psi(0) |^2}{m^2} \, 
\bigg \lbrace 
2 (1 - y_1) 
\left [
2 a (2 - y_1) - y_1 \ln \left | \frac{1 + a}{1 - a} \right |
\right ]
\nonumber \\
& + & [ y_1^2 + (1 - y_1)^2 ] \ln \left | \frac{b + a}{b - a} \right |
\bigg \rbrace ,
\label{eq:distry} \\
a & = &
\frac{\sqrt{y_1^2 - \mu^2}}{2 - y_1}, 
\qquad
b = \frac{2 y_1 (1 - y_1) + \mu^2}{2 (1 - y_1) (2 - y_1)}.
\nonumber
\end{eqnarray}

\begin{figure}[tb]
\begin{center}
\centerline{\epsfxsize=\textwidth \epsffile[150 445 490 645]{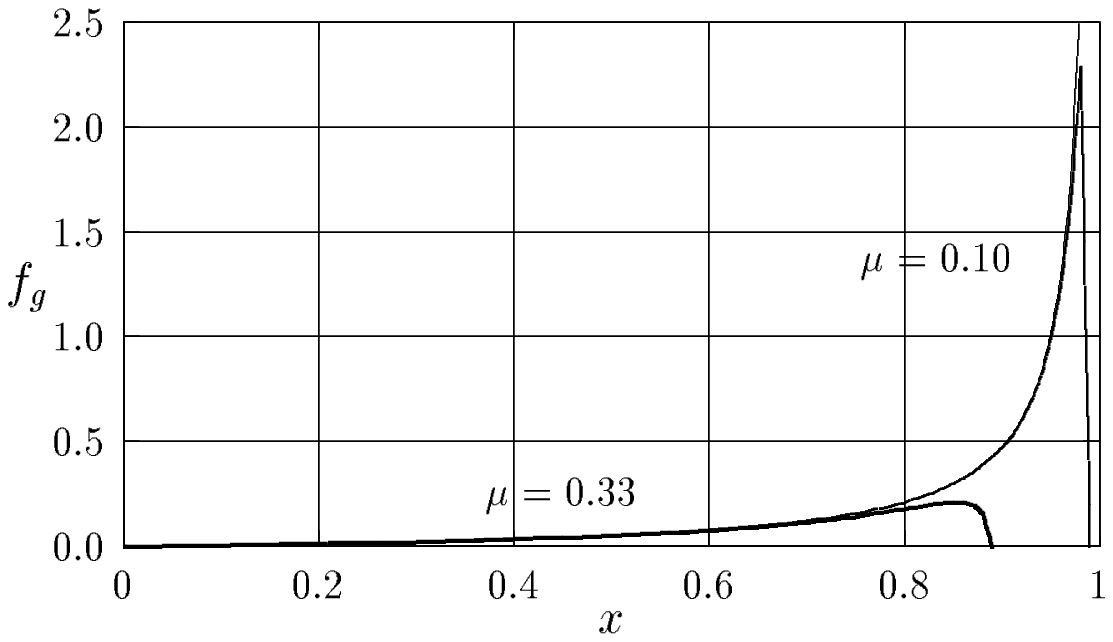}}
\vfill
A.Ya.~Parkhomenko and A.D.Smirnov \\ 
Mass effects in the quark-gluon decays of heavy paraquarkonia \\ 
\end{center}
\caption{}
\label{fig:disx}
\end{figure}

\begin{figure}[tb]
\begin{center}
\centerline{\epsfxsize=\textwidth \epsffile[150 445 490 645]{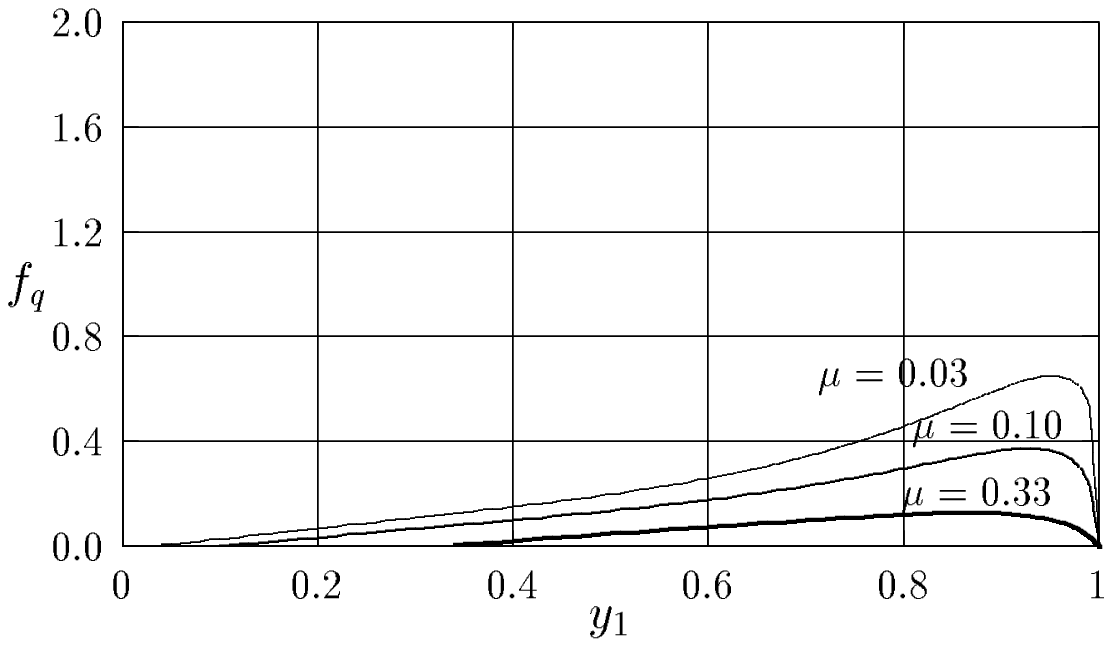}}
\vfill
A.Ya.~Parkhomenko and A.D.Smirnov \\ 
Mass effects in the quark-gluon decays of heavy paraquarkonia \\ 
\end{center}
\caption{}
\label{fig:disy}
\end{figure}

 For further analysis it is convenient to normalize
 these distribution functions as 
\begin{eqnarray}
f_g & = &\frac{1}{\alpha_s \, \Gamma_{2g}} 
      \frac{d\Gamma_{\bar q q g}}{dx},
\qquad
f_q = \frac{1}{\alpha_s \, \Gamma_{2g}} 
      \frac{d\Gamma_{\bar q q g}}{dy_1},
\nonumber
\end{eqnarray}
where
\begin{eqnarray}
\Gamma_{2g} \equiv \Gamma(^1S_0 (\bar Q Q) \to 2 g) = 
       F_c \frac{8 \pi \alpha_s^2}{m^2} | \psi (0) |^2
\nonumber
\end{eqnarray}
 is the two-gluon decay width of the paraquarkonium. 
 The normalized distribution functions $f_g$ and 
 $f_q$ in the gluon and quark energies are presented
 in Figs.~\ref{fig:disx} and~\ref{fig:disy} 
 at the relative final quark masses $\mu = 0.33$, 
 $\mu = 0.10$ and $\mu = 0.03$ corresponding to the decays 
 $\eta_b \to \bar c c g$, $\eta_c \to \bar s s g$ 
 and $\eta_b \to \bar s s g$ respectively. 
 
 As seen from Fig.~\ref{fig:disx} the gluon energy distribution 
 function $f_g$ has the strong enhancement
 at the end of the spectrum in the case of the small relative mass
 $\mu = 0.1$ ($\eta_c \to \bar s s g$)
 (as well as in the case of $\mu = 0.03$ ($\eta_b \to \bar s s g$)), 
 whereas this enhancement is absent in the 
 $\eta_b \to \bar c c g$ decay 
 because of the large relative mass of $c$-quark ($\mu = 0.33$).   
 The origin of this effect is the quark-antiquark collinear enhancement.

The quark energy distribution function $f_q$ (see Fig.~\ref{fig:disy}) 
has a small enhancement for $\mu = 0.10$ and $\mu = 0.03$ caused 
by an emission of the soft gluons. 

\begin{figure}[tb]
\begin{center}
\centerline{\epsfxsize=\textwidth \epsffile[150 445 490 645]{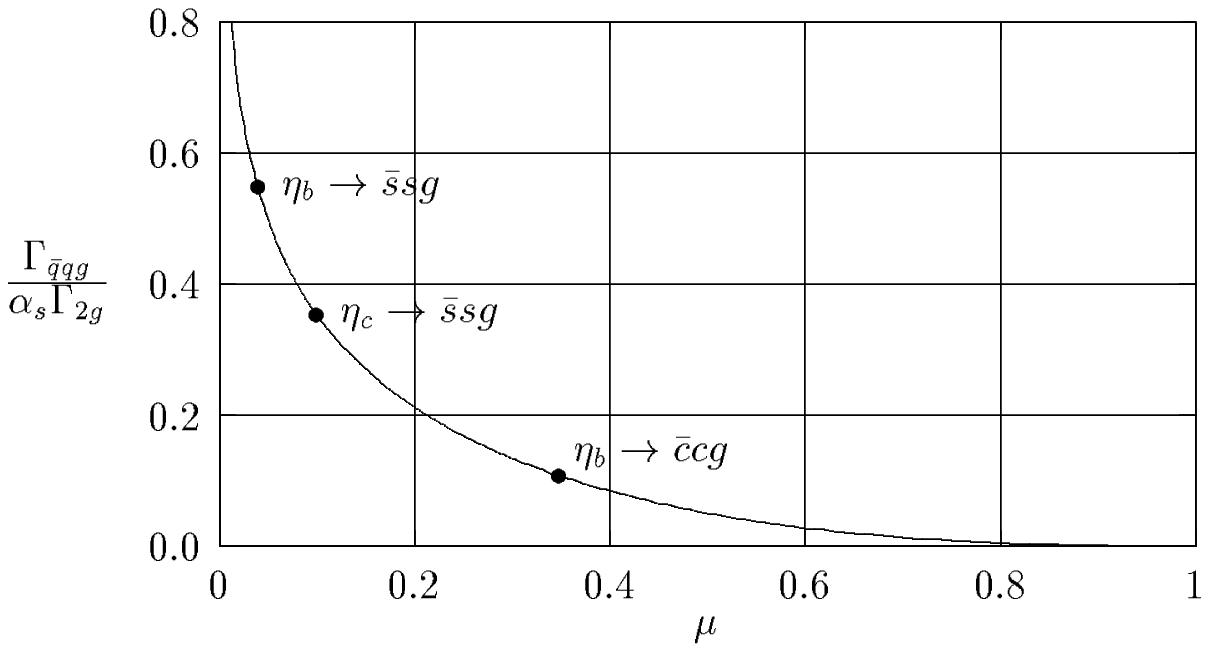}}
\vfill
A.Ya.~Parkhomenko and A.D.Smirnov \newline
Mass effects in the quark-gluon decays of heavy paraquarkonia \newline 
\end{center}
\caption{}
\label{fig:totwid}
\end{figure}

Integrating the distribution~(\ref{eq:distrx}) over the gluon energy $x$  
(or the distribution~(\ref{eq:distry}) over the quark energy $y_1$) we 
have obtained the total width $\Gamma_{\bar q q g}$ of the quark-gluon 
decay of the paraquarkonium with account of the masses of final quarks 
in the form: 

\begin{equation}
\Gamma_{\bar q q g} = 
F_c \, \frac{8 \alpha_s^3 | \psi(0) |^2}{3 m^2} \, 
\left \{
\ln \left | \frac{1 + \sqrt{1 - \mu^2}}{1 - \sqrt{1 - \mu^2}} \right | -
\frac{2}{3} \, (4 - \mu^2) \, \sqrt{1 - \mu^2}  
\right \} .
\label{eq:prob}
\end{equation} 
The dependence of this width on the masses of the final 
quarks can be described by the function 
 
\begin{eqnarray}
f_{\rm tot} = \frac{\Gamma_{\bar q q g}}{\alpha_s \Gamma_{2g}} =
\frac{1}{3 \pi} \, 
\left \{
\ln \left | \frac{1 + \sqrt{1 - \mu^2}}{1 - \sqrt{1 - \mu^2}} \right | -
\frac{2}{3} \, (4 - \mu^2) \, \sqrt{1 - \mu^2} 
\right \} ,
\label{eq:probdl}
\end{eqnarray}

\noindent 
 shown in Fig.~\ref{fig:totwid}.
 The points on the curve in Fig.~\ref{fig:totwid} correspond to  
 the $\eta_c$ meson decay $\eta_c \to \bar s s g$ and to the 
 $\eta_b$ meson decays $\eta_b \to \bar c c g$ and $\eta_b \to \bar s s g$. 
 As seen, the dependence of the width $\Gamma_{\bar q q g}$ 
 on the final quark masses is essential in analyzing the decays 
 of the $\eta_c$ and $\eta_b$ mesons. 

 For numerical estimations 
 we use further the values $\alpha_s (m_b) \approx 0.16$ and
 $\alpha_s (m_c) \approx 0.21$ resulting from 
 $\alpha_s (m_Z) = 0.118 \pm 0.003$~\cite{PDG}.
 Using the width~(\ref{eq:prob}) and the theoretical expression 
 for the hadronic width of the paraquarkonium
 $\Gamma_{\rm had} = \Gamma_{2g} (1 - C \alpha_s / \pi)$
 with the $C = 5.84$ for charmonia and 
 $C = 5.41$ for bottomonia~\cite{Barb1} 
 we have calculated the branching ratios 
 ${\rm Br} = \Gamma_{\bar q q g} / \Gamma_{\rm had}$
 of $^1S_0 (\bar Q Q) \to \bar q q g$ decays.
 The results are presented in Table~\ref{tab:predictions} for
 the final quark masses varying from their current values 
 (the first values of $m_q$ and Br) 
 to constituent ones (the last values of $m_q$ and Br).
 
 The result concerning the $\eta_c \to \bar s s g$
 decay can be compared with the experimental value of 
 ${\rm Br} (\eta_c \to K \bar K + X)$ (where $X$ is anything else) 
 presented in Table~\ref{tab:data}.
 The theoretical prediction of ${\rm Br} (\eta_c \to \bar s s g) = 12\%$ 
 at the current final quark masses is in good agreement with the
 experimental value ${\rm Br} (\eta_c \to K \bar K + X) = 11.6 \pm 2.4\%$,
 i.e. the summed width of the $\eta_c \to K \bar K + X$ type decays
 is saturated mainly by the subprocess $\eta_c \to \bar s s g$.
 One may hope that the branching ratios   
 ${\rm Br} (\eta_b \to \bar c c g) \approx 2.5\%$ and 
 ${\rm Br} (\eta_b \to \bar s s g) \approx 13\%$ 
 presented in Table~\ref{tab:predictions} also give a good approximation
 for the branching ratios of  
 the parabottomonium decays $\eta_b \to D \bar D + X$  
 and $\eta_b \to K \bar K + X$.

 We summarize the results of this letter as follows:  

 The quark-gluon decays of heavy paraquarkonia
 $^1 S_0 (\bar Q Q) \to \bar q q g$ are investigated in the 
 tree approximation taking account of the final quark masses. 

 The distribution functions in quark and gluon energies 
 are obtained and discussed. The distribution function in 
 the gluon energy is shown to have the strong collinear 
 enhancement in all the decays of $\eta_c$ and $\eta_b$ mesons 
 of $^1S_0 (\bar Q Q) \to \bar q q g$ type except the decay  
 $\eta_b \to \bar c c g$. 

 The total width of $^1S_0 (\bar Q Q) \to \bar q q g$ 
 decay is calculated and its dependence on the final quark masses 
 is shown to be essential for decays of $\eta_c$ and $\eta_b$ mesons. 
 The corresponding branching ratios of $\eta_c$ and $\eta_b$ mesons  
 are numerically estimated with a good agreement of the branching 
 ratio ${\rm Br} (\eta_c \to \bar s s g)$
 with experimental value of ${\rm Br} (\eta_c \to \bar K K + X)$. 

\bigskip

{\bf Acknowledgments}

\medskip

This work was partially supported by the programme 
``Universities of Russia -- Basic Research'' under grant No.~1157.
The work of A.Ya.~Par\-kho\-men\-ko was partially supported by the 
Russian Foundation for Basic Research under grant No.~98-02-16694.


\newpage

\begin{table}[htb]
\caption{Theoretical predictions for the branching ratios 
         ${\rm Br} = \Gamma_{\bar q q g} / \Gamma_{\rm had}$
         of the quark-gluon decays of heavy paraquarkonia.}
\label{tab:predictions}
\vspace*{10mm}
\begin{center}
\begin{tabular}{|c|c|c|}
\hline
Decay & $m_q$ (MeV) & Br (\%) \\
\hline
$\eta_c \to \bar s s g$ & $ 150 \div 450 $ & $ 12 \div 5 $ \\
$\eta_c \to \bar d d g$ & $ 7 \div 300   $ & $ 35 \div 7 $ \\
$\eta_c \to \bar u u g$ & $ 4 \div 300   $ & $ 38 \div 7 $ \\
\hline
$\eta_b \to \bar c c g$ & $ 1500 $         & $ 2.5 $ \\
$\eta_b \to \bar s s g$ & $ 150 \div 450 $ & $ 13 \div  8 $ \\
$\eta_b \to \bar d d g$ & $ 7 \div 300 $   & $ 27 \div 10 $ \\
$\eta_b \to \bar u u g$ & $ 4 \div 300 $   & $ 30 \div 10 $ \\
\hline
\end{tabular}
\end{center}
\end{table}
\vfill

\newpage

\begin{table}[htb]
\caption{Experimental values~\cite{PDG} of the fractions of the 
         $\eta_c \to K \bar K + X$ type decays ($X$ is anything else).}
\label{tab:data}
\vspace*{10mm}
\begin{center}
\begin{tabular}{|c|c|}
\hline
Modes & Fractions \\[3pt]
\hline
\raisebox{-2pt}{$K \bar K \pi$} & $ (5.5 \pm 1.7) \% $ \\
$K^{*0} K^- \pi^+ + c.c.$       & $ (2.0 \pm 0.7) \% $ \\
$K^+ K^- \pi^+ \pi^- $          & $ (2.0^{+ 0.7}_{-0.6}) \% $ \\
$ 2(K^+K^-)$                    & $ (2.1 \pm 1.2) \% $ \\  [3pt]
\hline
\raisebox{-2pt}{$K \bar K + X$} & $ (11.6 \pm 2.4) \% $ \\
\hline 
\end{tabular}
\end{center}
\end{table}
\vfill

\newpage

{\Large\bf Figure captions}

\bigskip

\begin{itemize}

\item[Fig. 1.] 
         Graphs describing the quark-gluon decay of a heavy paraquarkonium 
         $^1 S_0 (Q \bar Q) \to q \bar q g$.

\item[Fig. 2.] 
         Distribution functions $f_g$ in the gluon energy $x$
         at the relative masses of final quarks 
         $\mu = 0.33$ ($\eta_b \to \bar c c g$) and
         $\mu = 0.10$ ($\eta_c \to \bar s s g$).

 \item[Fig. 3.] 
         Distribution functions $f_q$ in the quark energy $y_1$
         at the relative masses of final quarks 
         $\mu = 0.33$ ($\eta_b \to \bar c c g$),
         $\mu = 0.10$ ($\eta_c \to \bar s s g$) and
         $\mu = 0.03$ ($\eta_b \to \bar s s g$).

\item[Fig. 4.] 
         The normalized width $f_{\rm tot}$ of the decay 
         $^1S_0 (\bar Q Q) \to \bar q q g$ as a function 
         of the relative final quark mass $\mu$. 

\end{itemize}

\newpage

%
%
%

\newpage

%
%

%
\vfill
%

%

\end{document}